\documentclass{article}

\usepackage{arxiv}
\usepackage[utf8]{inputenc}
\usepackage[T1]{fontenc}
\usepackage{hyperref}
\usepackage{url}
\usepackage{booktabs}
\usepackage{amsfonts}
\usepackage{nicefrac}
\usepackage{microtype}
\usepackage{graphicx}
\usepackage{natbib}
\usepackage{doi}
\usepackage{amsmath}

\title{Code2Doc: A Quality-First Curated Dataset for Code Documentation}

\date{}

\author{
	Recep Kaan Karaman \\
	Department of Computer Engineering\\
	Uludag University\\
	\texttt{kaankaraman@uludag.edu.tr} \\
	\And
	Meftun Akarsu \\
	M.Eng. AI Engineering of Autonomous Systems \\
	Technische Hochschule Ingolstadt \\
	\texttt{mea5963@thi.de} \\
}

\renewcommand{\shorttitle}{Code2Doc Dataset}

\hypersetup{
pdftitle={Code2Doc: A Quality-First Curated Dataset for Code Documentation},
pdfauthor={Recep Kaan Karaman, Meftun Akarsu},
pdfkeywords={Code Documentation, Dataset Curation, Software Engineering, Machine Learning, Quality Filtering},
}

\begin{document}
\maketitle

\begin{abstract}
The performance of automatic code documentation generation models depends critically on the quality of the training data used for supervision. However, most existing code documentation datasets are constructed through large scale scraping of public repositories with limited quality control. As a result, they often contain noisy documentation, extensive duplication, and increasing contamination from AI generated content. These issues weaken the supervision signal available to learning-based models and complicate evaluation.

We introduce \textbf{Code2Doc}, a quality-first curated dataset for function-level code documentation generation. Code2Doc consists of 13,358 high-quality function-documentation pairs extracted from widely used open-source repositories spanning five programming languages: Python, Java, TypeScript, JavaScript, and C++. The dataset is constructed using a four-stage curation pipeline that enforces documentation completeness and clarity, filters functions based on structural and complexity criteria, removes exact and near-duplicate code, and identifies documentation likely to be AI generated. Starting from 52,069 extracted candidates, only 25.6\% satisfy all quality constraints.

We provide a detailed analysis of the resulting dataset, which achieves a mean documentation quality score of 6.93 out of 10. Overall, 86.9\% of samples contain explicit type annotations, and only 2.9\% are flagged as potentially AI generated. Baseline experiments show that fine-tuning a large language model on Code2Doc yields relative improvements of 29.47\% in BLEU and 24.04\% in ROUGE-L over zero shot performance, despite the modest dataset size. We release both the dataset and the full curation pipeline to support reproducible research on automatic code documentation generation.
\end{abstract}


\section{Introduction}

High quality code documentation plays a central role in software maintenance, program comprehension, and collaborative development. At the function level, documentation serves as a compact description of intent and behavior, enabling developers to reason about code without inspecting implementations in detail. In practice, however, function-level documentation is often incomplete, inconsistent, or entirely missing, particularly in large and long-lived software systems.

Recent progress in large language models has led to renewed interest in automatic code documentation generation. While these models demonstrate strong generative capabilities, their performance depends heavily on the quality of the training data used for supervision. Most existing datasets for code-to-text generation are constructed through large scale scraping of public repositories with limited quality control. As a result, they frequently contain noisy or misaligned documentation, extensive duplication across projects, and documentation that does not accurately reflect the corresponding code.

These issues have become more pronounced in recent datasets due to the growing presence of AI generated documentation in open source repositories. Such content introduces additional challenges for both training and evaluation, as models may learn to reproduce templated or generic descriptions rather than project-specific semantics. Despite increasing awareness of data quality concerns in machine learning, existing code documentation datasets continue to prioritize scale over curation.

No publicly available dataset explicitly addresses documentation quality, deduplication, and machine generated content detection at the function level across multiple languages. Current resources either emphasize dataset size or focus on downstream benchmarks without addressing the quality of the underlying supervision signal.

In this work, we introduce Code2Doc, a curated dataset designed to support high quality research on automatic code documentation generation. Code2Doc is constructed using a four-stage filtering pipeline that enforces documentation completeness and clarity, filters functions based on structural and complexity criteria, removes exact and near-duplicate code, and identifies documentation that is likely to be AI generated. Starting from 52,069 extracted function–documentation pairs, only 25.6\% satisfy all quality constraints, highlighting the scarcity of consistently high quality documentation even in widely used open-source projects.

Our contributions are as follows. First, we present a transparent and reproducible curation pipeline for function-level code documentation datasets. Second, we release Code2Doc, a multilingual dataset of 13,358 high quality function–documentation pairs spanning five programming languages and accompanied by rich metadata. Third, we provide baseline experiments demonstrating that quality-focused curation enables effective model specialization despite substantially smaller dataset size

\section{Related Work}
\label{sec:related}

\subsection{Code Documentation Datasets}

The construction of large-scale datasets for code documentation generation has been driven primarily by advances in neural code-to-text models. \textbf{CodeSearchNet}~\cite{husain2019codesearchnet} pioneered this direction by scraping GitHub repositories across six programming languages, yielding millions of function-documentation pairs. While CodeSearchNet established important benchmarks, its reliance on minimal filtering has been documented to result in substantial noise, including misaligned documentation and auto-generated boilerplate~\cite{haque2022semantic}.

\textbf{CodeXGLUE}~\cite{DBLP:journals/corr/abs-2102-04664} extended this work by providing standardized evaluation protocols across multiple code intelligence tasks, including code summarization. However, CodeXGLUE inherits much of its underlying data from CodeSearchNet and similarly prioritizes benchmark construction over data quality assessment.

More recent large-scale corpora such as \textbf{The Stack}~\cite{thestack}, \textbf{StarCoder}~\cite{li2023starcodersourceyou}, and \textbf{The Stack v2}~\cite{lozhkov2024starcoder2stackv2} contain trillions of tokens of source code scraped from public repositories. These datasets support pretraining of large language models for code but do not provide curated function-documentation pairs or enforce alignment between code and natural language descriptions.

\textbf{CoDesc}~\cite{hasan2021codesclargecodedescriptionparallel} addresses data quality by applying semantic filtering techniques to improve alignment between code and documentation. However, CoDesc focuses exclusively on Python and does not address duplication or AI generated content. No existing publicly available dataset simultaneously addresses documentation quality assessment, aggressive deduplication, detection of machine-generated content, and multilingual coverage at the function level. Code2Doc fills this gap by prioritizing curation over scale.

\subsection{Data Quality and Deduplication}

The importance of data quality over dataset size has been increasingly recognized. \textbf{DataComp}~\cite{datacomp} demonstrated that carefully filtered datasets can outperform much larger unfiltered corpora on vision-language tasks. \textbf{LIMA}~\cite{lima} showed that alignment of large language models can be achieved with fewer than 1,000 carefully curated examples.

Several works have identified specific data quality issues in code datasets. \textbf{Duplication and memorization} have been shown to inflate model performance while reducing generalization~\cite{allamanis2019adverseeffectscodeduplication}. Studies of CodeSearchNet found that 15--25\% of test samples are near-duplicates of training data~\cite{allamanis2019adverseeffectscodeduplication}. \textbf{Dataset contamination} from large-scale web scraping has become a growing concern~\cite{golchin2023time}.

For deduplication, \textbf{The Stack}~\cite{thestack} applied MinHash and locality-sensitive hashing but focused on file-level rather than function-level granularity. Code2Doc adopts a similar two-phase strategy combining exact hash-based removal and MinHash-LSH approximate matching at the function level.

The emergence of AI generated code in public repositories introduces additional challenges. Recent 2025 industry reports estimate that 41\% of all code written in 2025 is AI generated~\cite{ai_generated_code_2025}, reflecting the deep integration of generative tools like Copilot and similar systems into developer workflows. Code2Doc addresses these concerns through aggressive deduplication, quality scoring, and heuristic detection of AI generated documentation.

\subsection{Code Complexity Metrics}

Cyclomatic complexity~\cite{mccabe1976complexity} remains the most widely used metric for quantifying code complexity based on control flow. It has been shown to correlate with defect density, maintenance effort, and comprehension difficulty. Recent work has explored learned complexity metrics using neural networks trained on developer annotations~\cite{scalabrino2019automatically}.

In Code2Doc, we use cyclomatic complexity as a filtering criterion to exclude both trivial functions and excessively complex functions unlikely to benefit from automated documentation generation.

\section{Methodology}
\label{sec:methodology}

\subsection{Repository Selection}

We curate documentation from widely used open-source repositories written in Python, Java, TypeScript, JavaScript, and C++. Repositories are selected based on adoption, maintenance activity, licensing, documentation practices, and domain diversity. Covered domains include web frameworks, machine learning libraries, developer tooling, distributed systems, compilers, and infrastructure software. In total, the dataset is curated from over 30 widely used repositories.

\subsection{Data Extraction}

We use Tree-sitter to extract function definitions and associated documentation. Language specific parsers traverse abstract syntax trees to identify functions, extract docstrings or documentation comments, record function signatures and type annotations, compute cyclomatic complexity, and collect metadata.

This process yields an initial pool of \textbf{52,069} function-documentation pairs.

\subsection{Four-Stage Filtering Pipeline}

\subsubsection{Stage 1: Basic Filtering}

We remove trivially documented or degenerate samples, including extremely short or excessively long documentation, very small or very large functions, test code, trivial accessors, and documentation containing placeholder text (e.g., \texttt{TODO}, \texttt{FIXME}). Exact filtering thresholds are summarized in Appendix~A (Table~\ref{tab:filtering_thresholds}).

\subsubsection{Stage 2: Quality Scoring}

Remaining samples are scored on a 0--10 scale using eight weighted dimensions capturing documentation completeness, parameter and return value coverage, use of type annotations, clarity, structural consistency, appropriate code complexity, and overall code quality. Only samples with a score $\geq 6.0$ are retained.

Formally, let $f$ denote a function--documentation pair. Each sample is evaluated along
$N = 8$ quality dimensions, including documentation completeness, parameter coverage,
return value description, use of type annotations, clarity, structural consistency,
appropriate code complexity, and overall code quality.

Let $q_i(f) \in [0,1]$ denote the normalized score of function $f$ on quality dimension $i$,
and let $w_i > 0$ be the corresponding weight assigned to that dimension. The overall
quality score $Q(f)$ is computed as a weighted average:

\[
Q(f) = \frac{1}{\sum_{i=1}^{N} w_i} \sum_{i=1}^{N} w_i \cdot q_i(f).
\]

The resulting score is linearly scaled to the range $[0,10]$. Only samples satisfying
$Q(f) \geq 6.0$ are retained for subsequent stages of the pipeline.

\subsubsection{Stage 3: Deduplication}

Despite prior filtering, large open-source repositories often contain duplicated or
near-duplicated functions due to shared utilities, code reuse, or minor refactoring.
To reduce redundancy and prevent over-representation of repeated patterns, we apply
a two-phase deduplication strategy.

\paragraph{Exact Duplicate Removal.}
Each function's source code is first normalized by removing comments and collapsing
whitespace. Let $\tilde{c}_f \in \Sigma^*$ denote the normalized code string of function $f$.
Two functions $f$ and $g$ are considered exact duplicates if:
\[
\tilde{c}_f = \tilde{c}_g.
\]
Exact duplicates are removed using hash-based lookup, retaining the first occurrence.

\paragraph{Token Set Representation.}
For near-duplicate detection, normalized code is tokenized into lowercase alphanumeric
identifiers. Each function $f$ is represented as a set of tokens:
\[
S_f = \{t_1, t_2, \dots, t_n\},
\]
where token multiplicity is ignored.

\paragraph{MinHash Signatures.}
To efficiently approximate pairwise similarity, we compute MinHash signatures using
$k$ independent hash functions $h_1, \dots, h_k : \mathcal{U} \rightarrow \mathbb{N}$.
The MinHash signature of function $f$ is defined as:
\[
\mathbf{m}(f) =
\left[
\min_{t \in S_f} h_1(t),
\min_{t \in S_f} h_2(t),
\dots,
\min_{t \in S_f} h_k(t)
\right].
\]
MinHash provides an unbiased estimator of Jaccard similarity between token sets.

\paragraph{Locality-Sensitive Hashing.} MinHash signatures are inserted into a locality-sensitive hashing (LSH) index configured with a similarity threshold $\tau$. For each function, we query the index for previously inserted functions exceeding this threshold. If such a function exists, the current sample is discarded as a near-duplicate; otherwise, it is retained and added to the index.

Deduplication is performed in a single deterministic pass, ensuring reproducibility without requiring explicit pairwise similarity computation.

\subsubsection{Stage 4: AI Generated Content Detection}

Recent open source repositories increasingly contain documentation generated or assisted by large language models. To reduce contamination and prevent feedback loops during model training, we apply a lightweight heuristic-based detector to identify documentation that is likely AI generated.

\paragraph{Heuristic Signals.}
Rather than training a supervised classifier, we adopt a rule-based approach that
assigns an \emph{AI likelihood score} based on the presence of multiple independent
signals commonly observed in AI generated documentation:

AI generated
\begin{itemize}
    \item \textbf{GPT-style phrase patterns:} templated instructional language and
    boilerplate phrasing frequently produced by large language models
    \item \textbf{Suspicious structural patterns:} unusually uniform formatting or
    repetitive section ordering
    \item \textbf{Perfect structural completeness:} simultaneous presence of nearly
    all standard documentation sections (e.g., description, parameters, returns,
    raises, examples)
    \item \textbf{Generic template-like language:} formulaic expressions that convey
    little project-specific or domain-specific information
\end{itemize}

\paragraph{Scoring Function.}
Let $d$ denote a documentation string. Each heuristic $h_j(d)$ contributes a fixed increment $\alpha_j$ to an aggregate confidence score. The AI likelihood score is computed as:
\[
C(d) = \min\left(1.0, \sum_{j} \alpha_j \cdot \mathbb{I}[h_j(d)] \right),
\]
where $\mathbb{I}[\cdot]$ is the indicator function. In our implementation, GPT-style phrase matches contribute $0.3$, suspicious structural patterns contribute $0.2$, perfect structural completeness contributes $0.2$, and generic language contributes $0.1$. The score is capped at $1.0$.

A documentation sample is flagged as potentially AI generated if:
\[
C(d) \geq \tau,
\]
where $\tau$ is a configurable threshold. We use a conservative threshold to favor precision over recall, ensuring that only high confidence cases are flagged.

\paragraph{Perfect Structure Detection.}
We identify suspiciously perfect structure by counting the presence of standard documentation sections. If at least four of the following sections are present, the documentation is flagged as structurally suspicious: \emph{Description}, \emph{Parameters/Args}, \emph{Returns}, \emph{Raises/Throws}, and \emph{Examples}.

\paragraph{Usage and Metadata.}
For each sample, the AI likelihood score and matched heuristic patterns are stored as metadata. Depending on configuration, flagged samples may either be removed or retained with a manual review flag. In this work, flagged samples are retained in the dataset but explicitly marked.

\paragraph{Impact.}
Across the final dataset, 387 samples (2.9\%) are flagged as potentially AI generated. This relatively low rate supports the effectiveness of temporal filtering and the conservative nature of our heuristic detection strategy.

\section{Dataset Analysis}

\begin{figure}[t]
\centering
\includegraphics[width=0.85\linewidth]{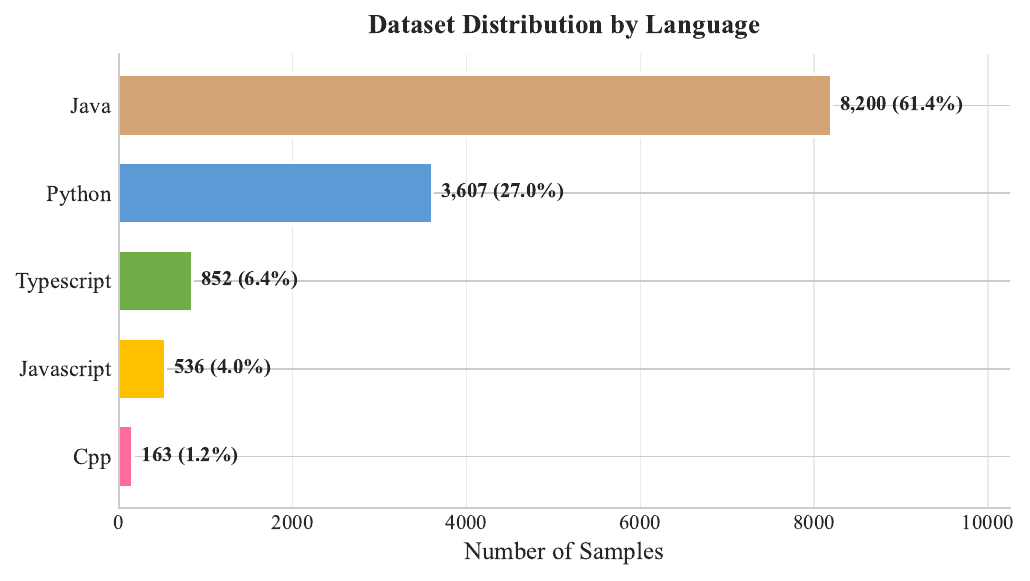}
\caption{Distribution of samples across programming languages in the Code2Doc dataset.
Java and Python dominate due to stronger documentation practices in mature enterprise
and scientific software ecosystems.}
\label{fig:language_dist}
\end{figure}

\begin{figure}[t]
\centering
\includegraphics[width=0.9\linewidth]{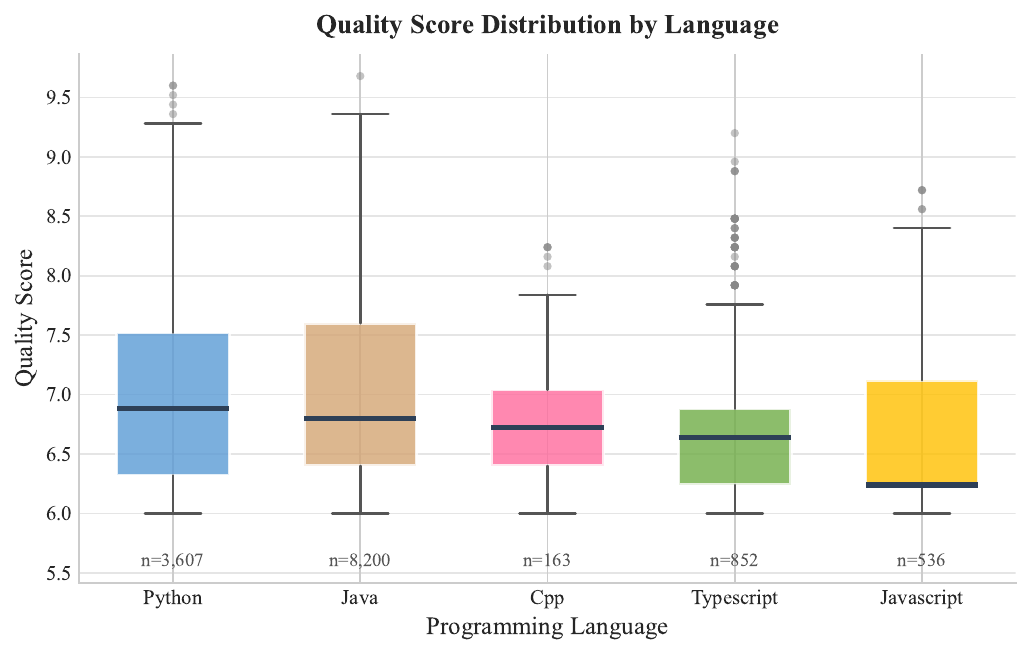}
\caption{Documentation quality scores grouped by programming language.
Scores are tightly distributed across languages, indicating consistent
quality enforcement despite differences in language ecosystems.}
\label{fig:quality_lang}
\end{figure}

\begin{figure}[t]
\centering
\includegraphics[width=0.85\linewidth]{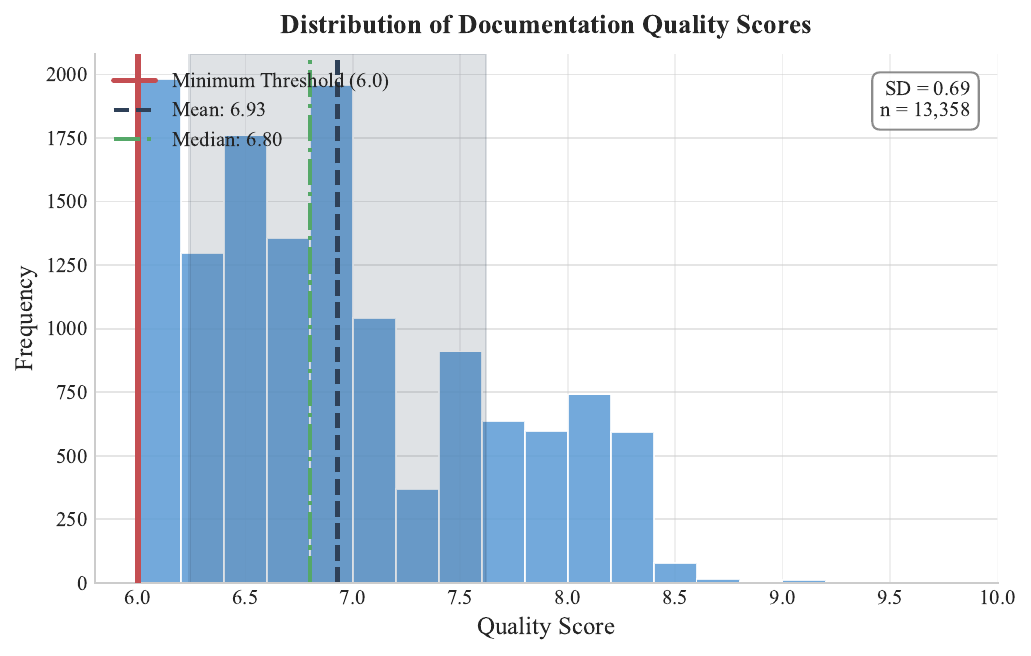}
\caption{Distribution of documentation quality scores in Code2Doc.
All retained samples exceed the minimum quality threshold of 6.0, with
a compact distribution centered around a mean of 6.93.}
\label{fig:quality_dist}
\end{figure}

In addition to aggregate statistics, we analyze the distribution of samples across programming languages and repositories, as well as the distribution of documentation quality scores. Figure~\ref{fig:language_dist} summarizes the language composition of Code2Doc, while Figures~\ref{fig:quality_lang} and~\ref{fig:quality_dist} provide insight into documentation quality across languages and overall.

\label{sec:dataset}

\begin{table}[t]
\caption{Summary statistics of the Code2Doc dataset.}
\label{tab:statistics}
\centering
\begin{tabular}{lr}
\toprule
\textbf{Metric} & \textbf{Value} \\
\midrule
Total samples & 13{,}358 \\
Mean quality score & 6.93 / 10 \\
Median quality score & 6.80 / 10 \\
Standard deviation & 0.69 \\
Score range & [6.00, 9.68] \\
\midrule
Mean cyclomatic complexity & 3.40 \\
Median cyclomatic complexity & 2.0 \\
\midrule
Samples with type annotations & 11{,}602 (86.9\%) \\
AI flagged samples & 387 (2.9\%) \\
\bottomrule
\end{tabular}
\end{table}

\subsection{Language Distribution}

The Code2Doc dataset spans five programming languages and reflects the distribution of high-quality documentation in mature open-source ecosystems. Java constitutes the majority of the dataset (61.4\%, 8{,}200 samples), followed by Python (27.0\%, 3{,}607 samples). TypeScript (6.4\%), JavaScript (4.0\%), and C++ (1.2\%) account for the remaining portion.

This imbalance mirrors real world documentation practices, where large, enterprise and infrastructure oriented Java codebases exhibit more consistent function-level documentation. All dataset splits are stratified by programming language to preserve this distribution during training and evaluation.

\subsection{Repository Coverage}

Code2Doc is curated from over 30 widely used open-source repositories spanning diverse domains, including web frameworks, machine learning libraries, distributed systems, compilers, and developer tooling. Large and mature projects such as \texttt{apache/commons-lang}, \texttt{spring-framework}, \texttt{spring-boot}, \texttt{apache/kafka}, \texttt{google/guava}, and \texttt{pandas} contribute a substantial fraction of the retained samples.

The resulting repository distribution exhibits a long-tailed structure. While some projects contribute over a thousand high-quality samples, many others yield only a small number after filtering. This outcome reflects the strict curation criteria applied during dataset construction and highlights the scarcity of consistently high-quality documentation even in widely adopted software projects.

\subsection{Documentation Quality}

Documentation quality scores are tightly distributed, with a mean of 6.93 and a standard deviation of 0.69. The enforced minimum threshold of 6.0 ensures that all samples meet a baseline level of completeness and clarity, while the upper tail of the distribution captures exceptionally well-documented functions.

The compactness of the quality distribution indicates that Code2Doc avoids extreme outliers and provides a relatively homogeneous supervision signal compared to large scale scraped datasets, which often contain highly variable documentation quality.

\subsection{Code Complexity}

Function complexity is intentionally constrained to focus on realistic and maintainable code. The median cyclomatic complexity is 2.0, and the mean is 3.40, indicating that most functions contain limited branching logic. This design choice avoids trivial accessors while excluding unusually complex functions that are unlikely to benefit from automated documentation generation.

\subsection{Type Annotations and Documentation Style}

A total of 86.9\% of samples include explicit type annotations in function signatures or documentation. This high prevalence is driven by the dominance of statically typed languages in the dataset and by modern Python projects that adopt gradual typing. Type information provides an additional supervision signal that is valuable for documentation generation models.

No samples in the final dataset contain executable documentation examples. This is a direct consequence of conservative filtering: examples in real-world repositories are frequently incomplete, outdated, or incorrect, and were therefore excluded to avoid introducing misleading supervision. Incorporating high-quality executable examples remains an avenue for future dataset extensions.

\section{Baseline Experiments}
\label{sec:experiments}

To demonstrate the utility of Code2Doc, we evaluate its effectiveness for training automatic documentation generation models.

\subsection{Experimental Setup}

We fine-tune Llama~3.1~8B~\cite{llama31} on the task of function-level documentation generation using parameter-efficient fine-tuning (PEFT).

\paragraph{Training Data.}
The full Code2Doc dataset consists of 13{,}358 samples spanning five programming languages: Python, Java, TypeScript, JavaScript, and C++. We consider two experimental settings (i) zero-shot evaluation using the base model without fine-tuning, and (ii) parameter-efficient fine-tuning on Code2Doc.

Fine-tuning is performed on the \emph{full multilingual training split}, exposing the model to documentation patterns across all five programming languages. No language specific subsets are used during training, allowing a single model to learn shared structural and stylistic regularities across languages.

\paragraph{Data Splits and Evaluation Protocol.}
The dataset is split into training, validation, and test sets using an 80/10/10 ratio, stratified by programming language to preserve the overall language distribution across splits. Evaluation is conducted on the held-out test set.

\paragraph{Fine-Tuning Configuration.}
Fine-tuning is performed using QLoRA~\cite{qlora} with LoRA adapters~\cite{lora}. Adapters are applied to both attention and feed-forward layers with rank $r=16$.

Training is conducted for three epochs using the AdamW \cite{loshchilov2019decoupledweightdecayregularization} optimizer in 8-bit precision with a learning rate of $1\times10^{-4}$ and a linear learning rate schedule with 50 warmup steps. The effective batch size is 8, achieved via a per-device batch size of 2 and gradient accumulation over 4 steps. Sequence packing is enabled to improve computational efficiency, and the maximum sequence length is set to 2048 tokens.

Flash Attention is leveraged through Unsloth’s optimized kernels when supported by the underlying hardware and configuration, reducing memory overhead and improving throughput for long sequences.

\begin{table}[t]
\centering
\caption{Fine-tuning configuration and hyperparameters used for training on Code2Doc.}
\label{tab:hyperparameters}
\begin{tabular}{ll}
\toprule
\textbf{Category} & \textbf{Setting} \\
\midrule
\multicolumn{2}{l}{\emph{Model and Adaptation}} \\
Base model & Llama 3.1 8B \\
Quantization & 4-bit (QLoRA) \\
LoRA rank ($r$) & 16 \\
LoRA scaling ($\alpha$) & 16 \\
LoRA dropout & 0 \\
Target modules & Q, K, V, O, gate, up, down projections \\
Gradient checkpointing & Enabled (Unsloth) \\
\midrule
\multicolumn{2}{l}{\emph{Training Configuration}} \\
Optimizer & AdamW (8-bit) \\
Learning rate & $1\times10^{-4}$ \\
Learning rate schedule & Linear \\
Warmup steps & 50 \\
Training epochs & 3 \\
Per-device batch size & 2 \\
Gradient accumulation steps & 4 \\
Effective batch size & 8 \\
Maximum sequence length & 2048 \\
Sequence packing & Enabled \\
\midrule
\multicolumn{2}{l}{\emph{Infrastructure and Reproducibility}} \\
Hardware & Single NVIDIA A100 (80GB) \\
Training platform & Google Colab Pro \\
Attention optimization & Unsloth optimized kernels (Flash Attention when supported) \\
Random seed & 3407 \\
\bottomrule
\end{tabular}
\end{table}

\subsection{Evaluation Metrics}

We evaluate generated documentation using a set of automatic metrics that are widely used in prior work on code summarization and documentation generation. Specifically, we report BLEU~\cite{papineni2002bleu} and ROUGE-L~\cite{lin2004rouge} which capture complementary aspects of generation quality.

BLEU measures n-gram overlap between a generated docstring and a reference docstring and provides an estimate of lexical similarity. Formally, BLEU computes a geometric mean of modified n-gram precisions with a brevity penalty to discourage overly short outputs. ROUGE-L is based on the length of the longest common subsequence between generated and reference texts, making it more sensitive to content coverage and sentence level structure.

Metrics are computed at the function level by comparing generated docstrings against ground truth documentation. Results are reported separately for each programming language and averaged over the corresponding test subsets.

While these metrics do not fully capture properties such as factual correctness, clarity, or developer usefulness, they provide a standardized basis for quantitative comparison and enable reproducible benchmarking against prior work.

\subsection{Results}

Table~\ref{tab:results} reports documentation generation performance on the Code2Doc benchmark for the base Llama~3.1~8B model in a zero-shot setting and after parameter-efficient fine-tuning on Code2Doc.

\begin{table}[t]
\centering
\caption{Documentation generation results on the Code2Doc benchmark.
Fine-tuning on Code2Doc yields consistent improvements over zero-shot performance.
Relative improvements with respect to the zero-shot baseline are shown in parentheses.}
\label{tab:results}
\begin{tabular}{lcc}
\toprule
\textbf{Model} & \textbf{BLEU} & \textbf{ROUGE-L} \\
\midrule
Zero-shot Llama~3.1~8B & 0.0302 & 0.0786 \\
Fine-tuned on Code2Doc & 0.0391 (+29.47\%) & 0.0975 (+24.04\%) \\
\bottomrule
\end{tabular}
\end{table}

Fine-tuning on Code2Doc improves BLEU from 0.0302 to 0.0391, corresponding to a relative gain of 29\%, and improves ROUGE-L from 0.0786 to 0.0975, corresponding to an 24\% relative improvement. These results indicate that quality-focused curation provides a meaningful supervision signal for documentation generation, despite the modest size of the dataset.

The observed gains suggest that reducing noise, duplication, and weakly aligned documentation enables more effective model specialization than reliance on large scale scraped corpora alone.

\section{Discussion}

Our results demonstrate that quality-focused curation enables effective model specialization despite substantially reduced dataset size. The 29\% relative improvement in BLEU and 24\% relative improvement in ROUGE-L achieved through fine-tuning on Code2Doc suggest that filtering noisy, duplicated, and misaligned samples provides a stronger supervision signal than relying on scale alone. This finding aligns with recent work in vision-language learning~\cite{datacomp} and instruction tuning~\cite{lima}, which similarly demonstrate that careful curation can match or exceed the performance of much larger unfiltered datasets.

\subsection{Quality Versus Scale Trade-offs}

The aggressive filtering applied during Code2Doc construction reduced the initial candidate pool from 52,069 samples to 13,358 retained samples, accepting only 25.6\% of candidates. This low retention rate highlights the scarcity of consistently high-quality documentation even in mature, widely used open-source projects. While this trade-off sacrifices dataset size, the resulting homogeneity in quality scores (mean 6.93, standard deviation 0.69) provides a more reliable foundation for both training and evaluation.

In contrast, large-scale datasets such as CodeSearchNet contain millions of samples but exhibit highly variable quality. Our results suggest that the supervision signal provided by such datasets may be weakened by low-quality examples, requiring models to implicitly learn to distinguish signal from noise during training. By performing this filtering explicitly, Code2Doc reduces the burden on the model and enables more efficient learning.

The modest absolute BLEU and ROUGE-L scores observed in our experiments (0.0391 and 0.0975, respectively) reflect the inherent difficulty of documentation generation rather than limitations of the dataset. Documentation quality cannot be fully captured by n-gram overlap metrics, which penalize semantically correct paraphrases and fail to account for factual accuracy or developer usefulness. Future work should incorporate human evaluation and task-based assessments to provide a more complete picture of model performance.

\subsection{Multilingual Composition and Transfer Learning}

Code2Doc spans five programming languages with a distribution that reflects real-world documentation practices. The dominance of Java (61.4\%) and Python (27.0\%) mirrors the prevalence of explicit function-level documentation in statically typed, enterprise-oriented projects and well-maintained scientific libraries. While this imbalance limits coverage of less-documented languages, it provides an authentic representation of where high-quality documentation currently exists.

The stratified language distribution across splits enables future research on cross-lingual transfer learning for documentation generation. Recent work has shown that code representations learned from multilingual pretraining can transfer across programming languages~\cite{wang-etal-2021-codet5, guo2021graphcodebert}, suggesting that documentation patterns may similarly transfer.

\subsection{AI Generated Content and Dataset Hygiene}

The detection and flagging of potentially AI generated documentation represents a novel contribution of Code2Doc. While only 2.9\% of samples were flagged, this conservative estimate reflects both temporal filtering and the precision oriented nature of our heuristics. The growing prevalence of AI assisted code in public repositories poses significant challenges for future dataset construction. Feedback loops, in which models are trained on data including outputs from previous generations, risk reducing diversity and amplifying biases~\cite{shumailov2023curse}.

Our heuristic-based approach identifies documentation exhibiting stylistic patterns characteristic of large language models, including GPT-style phrasing, suspiciously uniform structure, and generic template language. While this method cannot achieve perfect accuracy, it provides a practical mechanism for flagging samples that warrant manual review. Future iterations may benefit from supervised classifiers or zero-shot methods such as DetectGPT~\cite{mitchell2023detectgpt}.

\subsection{Implications for Software Engineering Practice}

Beyond training machine learning models, Code2Doc provides insights into documentation practices in real-world projects. The low retention rate during filtering (25.6\%) suggests that even in widely used, mature projects, the majority of function-level documentation does not meet basic quality standards. This underscores the need for better tooling, developer education, and incentive structures to improve documentation quality.

The dataset's focus on non-trivial functions (median cyclomatic complexity 2.0, mean 3.40) reflects code where documentation provides the greatest value. The absence of executable examples in the final dataset reflects the difficulty of maintaining accurate examples in real world repositories, most examples in scraped documentation were incomplete, outdated, or incorrect. This highlights a broader challenge: ensuring that documentation remains synchronized with evolving code.

\subsection{Limitations of Automatic Metrics}

While BLEU and ROUGE-L provide standardized benchmarks, they capture only surface-level lexical similarity and do not measure semantic correctness or developer usefulness. Recent work has proposed alternative metrics such as \textbf{CodeBERTScore}~\cite{zhou2023codebertscore}, which measures semantic similarity using learned representations, and task-based evaluations that assess whether documentation enables developers to correctly use code~\cite{mcburney2014automatic}. Future evaluations should incorporate these methods alongside human assessments.

\section{Limitations}

\subsection{Dataset Size and Coverage}

Code2Doc contains 13,358 samples, substantially smaller than large scale corpora such as CodeSearchNet (2.3 million samples). This tradeoff is intentional, but the modest size may limit utility for pretraining large models or covering rare programming patterns. Language coverage is also uneven: Java dominates (61.4\%), followed by Python (27.0\%), while TypeScript (6.4\%), JavaScript (4.0\%), and C++ (1.2\%) contribute smaller fractions. This imbalance reflects real-world documentation practices but may limit effectiveness for training language-specific models.

\subsection{Repository and Domain Bias}

Code2Doc is curated from widely used, mature open-source projects with strong documentation cultures, biasing the dataset toward enterprise software and infrastructure projects. Documentation patterns in less formal contexts—personal projects, research prototypes, or early stage startups are underrepresented. Similarly, the dataset focuses on general-purpose software development and does not extensively cover domain-specific contexts such as embedded systems or specialized application areas.

Geographically, the dataset likely reflects biases toward North American and European open-source communities, which dominate platforms like GitHub. Documentation styles and conventions from other regions may not be adequately represented.

\subsection{Quality Scoring and Human Validation}

The quality scoring mechanism is automated and based on heuristic features. While carefully designed to reflect documentation completeness, clarity, and alignment, it remains an approximation of true quality as perceived by developers. No inter-annotator reliability study has been conducted to validate alignment between automated scores and human judgments. The weights assigned to quality dimensions are chosen based on our assessment but have not been empirically validated or optimized.

\subsection{AI Generated Content Detection}

Our heuristic based approach to detecting AI generated documentation prioritizes precision over recall, likely missing AI generated content that does not exhibit specific stylistic patterns. Conversely, it may produce false positives, flagging high quality human written documentation matching our targeted patterns. Without ground truth labels or manual validation, we cannot precisely quantify false positive or false negative rates. As code generation tools become more sophisticated, rule-based detection effectiveness will likely degrade.

\subsection{Evaluation and Benchmarking}

Our baseline experiments rely on BLEU and ROUGE-L, which measure surface-level similarity but do not capture semantic correctness or developer usefulness. These metrics are known to poorly correlate with human judgments of documentation quality~\cite{reiter2018structured}. We do not conduct human evaluation of generated documentation, limiting our ability to assess whether improvements in automatic metrics translate to quality as perceived by developers. Task based evaluations would provide more ecologically valid assessments but were beyond the scope of this work.

\subsection{Temporal Validity}

Code2Doc is constructed from a snapshot of repositories and does not capture documentation evolution over time. As projects evolve, documentation may become outdated or misaligned with implementations. Our pipeline does not account for version history or temporal drift, which may introduce subtle misalignments. Additionally, the dataset does not capture recent trends in documentation practices or the increasing prevalence of hybrid human-AI authorship.

\section{Conclusion}

We presented Code2Doc, a curated dataset of 13,358 high quality function-documentation pairs across five programming languages. By prioritizing data quality over scale, Code2Doc provides a reliable foundation for training and evaluating code documentation generation models.

\appendix
\section{Filtering Thresholds and Design Choices}

\begin{table}[ht]
\centering
\caption{Thresholds and heuristics used during Stage~1 (Basic Filtering).}
\label{tab:filtering_thresholds}
\begin{tabular}{p{3.5cm} p{4.5cm} p{5.5cm}}
\toprule
\textbf{Criterion} & \textbf{Threshold / Heuristic} & \textbf{Rationale} \\
\midrule
Minimum docstring length & 20 characters &
Removes trivial or placeholder comments \\
Maximum docstring length & 10{,}000 characters &
Excludes anomalously verbose documentation \\
\midrule
Minimum cyclomatic complexity & 1 &
Filters out no-op or degenerate functions \\
Maximum reasonable complexity & 50 &
Excludes unusually complex, non-representative code \\
\midrule
Minimum logical lines & 5 &
Removes trivial wrappers and accessors \\
Accessor name prefixes &
\texttt{get}, \texttt{set}, \texttt{is}, \texttt{has} &
Identifies trivial getter/setter functions \\
\midrule
Test function patterns &
\texttt{test\_}, \texttt{test}, \texttt{Test}, \texttt{TEST}, \texttt{\_test} &
Excludes test-specific documentation \\
\midrule
Decision keywords &
Conditional and loop constructs &
Proxy for complexity estimation when needed \\
\bottomrule
\end{tabular}
\end{table}

All threshold values are centralized in the codebase and released as part of the dataset curation pipeline, which is archived on Zenodo to ensure reproducibility \cite{karaman_code2doc_pipeline_2025}. These thresholds are designed to remove degenerate or non-informative samples while preserving documentation patterns representative of real-world software systems.

\section{Data Availability}

The Code2Doc dataset and the full data curation pipeline are publicly available to
support reproducibility and future research.

\begin{itemize}
    \item \textbf{Dataset:} The curated Code2Doc dataset is hosted on Hugging Face at
    \url{https://huggingface.co/datasets/kaanrkaraman/code2doc}.
    \item \textbf{Curation and Training Code:} The complete preprocessing, filtering, and fine-tuning pipeline is provided as an executable Jupyter notebook at \url{https://colab.research.google.com/gist/kaanrkaraman/ceb9e4cbbc7757bc44bf554c4000f63b/llama-3-1-8b.ipynb}.
\end{itemize}

All artifacts are released under permissive open-source licenses and include
versioned metadata to ensure reproducibility.

\bibliographystyle{unsrt}
\bibliography{references}

\end{document}